\documentclass[showpacs,preprintnumbers,pre]{revtex4}
\usepackage{graphics,graphicx}
\usepackage{psfig,color}

\begin{document}
\title{The Vlasov equation and the Hamiltonian Mean-Field model
}
\author{Julien Barr{\'e}$^1$\thanks{E-mail: jbarre@math.unice.fr},
Freddy Bouchet$^{2}$\thanks{E-mail: Freddy.Bouchet@inln.cnrs.fr},
Thierry Dauxois$^3$\thanks{E-mail: Thierry.Dauxois@ens-lyon.fr},
Stefano Ruffo$^4$\thanks{E-mail: stefano.ruffo@unifi.it},
Yoshiyuki Y. Yamaguchi$^{5}$
\thanks{E-mail: yyama@amp.i.kyoto-u.ac.jp}}
\affiliation{
1. Laboratoire J.-A. Dieudonn\'e, UMR-CNRS 6621, Universit\'e de Nice, Parc Valrose, 06108 Nice cedex 02, France.\\
2. Institut Non Lin\'eaire de Nice,  UMR-CNRS 6618, 1361 route des Lucioles 06560 Valbonne, France.\\
3. Laboratoire de Physique, UMR-CNRS 5672, ENS Lyon, 46 All\'{e}e d'Italie, 69364 Lyon cedex 07, France\\
4. Dipartimento di Energetica ``S. Stecco" and CSDC, Universit{\`a} di Firenze, INFM and INFN, Via S. Marta, 3 I-50139, Firenze, Italy\\
5. Department of Applied Mathematics and Physics, Graduate School of Informatics, Kyoto University, 606-8501, Kyoto, Japan}

\date{\today}

\bigskip

\begin{abstract}
We show that the quasi-stationary states observed in the $N$-particle
dynamics of the Hamiltonian Mean-Field (HMF) model are nothing but
Vlasov stable homogeneous (zero magnetization) states. There is an infinity of Vlasov
stable homogeneous states corresponding to different initial
momentum distributions. Tsallis $q$-exponentials in momentum,
homogeneous in angle, distribution functions are possible,
however, they are not special in any respect, among an infinity of
others. All Vlasov stable homogeneous states lose their stability
because of finite $N$ effects and, after a relaxation time
diverging with a power-law of the number of particles, the system
converges to the Boltzmann-Gibbs equilibrium.
\end{abstract}


\maketitle

\section{Intriguing numerical results}

The Hamiltonian Mean-Field model (HMF)~\cite{antoni-95}
\begin{equation}
H=\displaystyle
\sum_{i=1}^{N}\frac{p_i^2}{2} +
\frac{1}{2N}\sum_{i,j=1}^{N} [1-\cos(\theta_i-\theta_j)],
\label{hamilHMF}
\end{equation}
describes the motion of globally coupled particles on a circle:
$\theta_i$ refers to the angle of the $i$-th particle and $p_i$
to its conjugate momentum, while $N$ is the total number of
particles. The $1/N$ prefactor, which has been historically
introduced to obtain an extensive energy, can be absorbed
in a time rescaling (we shall however keep it to compare
with previous results).

From the fundamental point of view, this is an ideal
toy model. Indeed, although it is simple and the mean-field
interaction allows us to perform analytical calculations, it has
several features of long-range interactions.
Moreover, it is a simplification of physical systems like
charged or gravitational sheet models.
Finally, in some cases, wave-particle Hamiltonians can be reduced
to it. In particular, for what the equilibrium properties are
concerned, the HMF Hamiltonian~(\ref{hamilHMF}) can be mapped
onto the Colson-Bonifacio model of the single-pass
Free Electron Laser~\cite{bbdrjstatphys}.

In this short note, we would like to emphasize that several
interesting numerical facts which have been reported in the literature
can be accurately explained by considering the limit of infinite number
of particles, namely the Vlasov equation corresponding to the
HMF model.
\begin{figure}[ht]
\begin{center}
\includegraphics[height=6truecm]{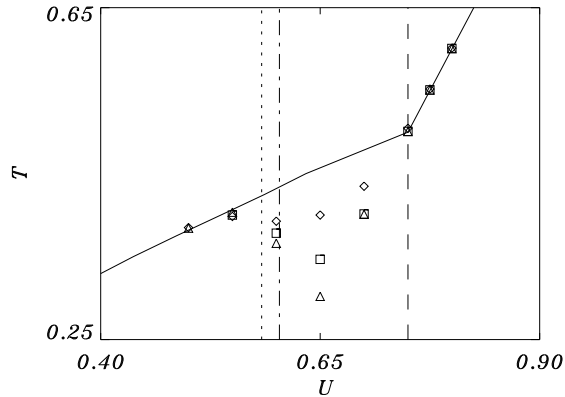}
\end{center}
\caption{Caloric curve of the HMF Hamiltonian. The solid line is
  the equilibrium result in both the canonical and the microcanonical
  ensemble. The second order phase transition is revealed by the kink
  at $U_c=3/4$. The three values of the energy indicated by the
  vertical lines are the stability thresholds for the homogeneous
  Gaussian (dashed), power-law of Eq.~(\ref{eq:powertail}) with
  $\nu=8$ (dash-dotted) and water-bag (dotted) initial momentum
  distribution. The Gaussian stability threshold coincides with the
  phase transition energy. The points are the results of constant
  energy (microcanonical) simulations for the Gaussian (losanges), the
  power-law (squares) and the water-bag (triangles). Simulations were
  performed with $N=5000$.  The tendency of the simulation points to
  lie on the continuation to lower energies of the supercritical
  branch of the caloric curve would increase when increasing the value
  of $N$. }
\label{figcaloric}
\end{figure}
The first numerical fact is the strong disagreement which was reported
in~Refs.~\cite{antoni-95,lrt2001} between constant energy molecular
dynamics simulations and canonical statistical mechanics calculations.
This unexpected and striking result, found for energies slightly below
the second order phase transition energy (see Fig.~\ref{figcaloric}),
was first thought to be the fingerprint of inequivalence between
microcanonical and canonical ensemble. It was known that such
inequivalence might have been present because of the long-range nature
of the interaction.  However, it has been later proved that
inequivalence occurs only if the system has a first order canonical
phase transition, which is not the case for the HMF, which has instead
a second order phase transition.  Moreover, the microcanonical entropy
of the HMF model has been recently derived using large deviation
theory~\cite{bbdrjstatphys}, showing that the two ensembles give the
same predictions.

\begin{figure}[ht]
\begin{center}
\includegraphics[height=6truecm]{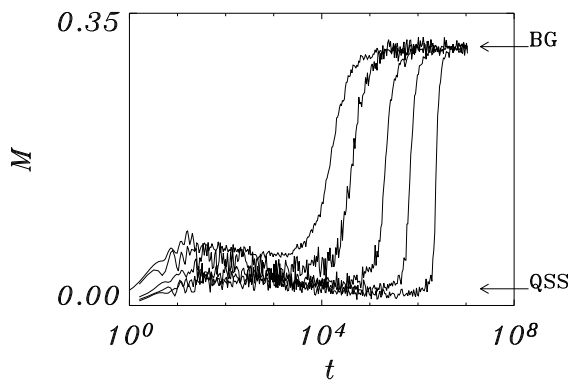}
\end{center} \caption{Time evolution of the modulus of the
magnetization $M(t)$ for different particle numbers: $N=10^3$,
$2.10^3$, $5.10^3$, $10^4$ and $2.10^4$ from left to right (U=0.69). In all
cases an average over several samples has been taken. Two values
of the magnetization, indicated by horizontal arrows, can be
identified in this figure: the upper one (labelled BG) corresponds
to the expected equilibrium result for the magnetization, while
the lower one, labelled QSS, represents the value of $M$ in the
quasi-stationary state.} \label{figdiffN}
\end{figure}

It then became clear that the disagreement must have a
{\em dynamical} origin. In order to characterize the dynamical
properties of the HMF model, the behaviour
of the modulus $M$ of the magnetization
\begin{equation}
{\bf M}=\displaystyle \frac{1}{N}\sum_ne^{i\theta_n}
\label{magnetization}
\end{equation}
has been typically studied. Its time evolution is shown on a
logarithmic scale in Fig.~\ref{figdiffN} for increasing values of
$N$. The initial state is a homogeneous in angle, zero
magnetization, state with water-bag distribution of momenta (see
next Section). The figure shows that the system evolves on a fast
timescale towards a state which has an almost zero magnetization
($M<0.05$ in our simulations). Such a state lasts for a long time
and its lifetime increases very rapidly with $N$ (note the
logarithmic timescale on the abscissa). States with such a
property have been called in the literature quasi-stationary
states (QSS). It is only in a second stage that the value of $M$
takes off and reaches the Boltzmann-Gibbs value (indicated by BG
in Fig.~\ref{figdiffN}) predicted by equilibrium statistical
mechanics.
This numerical result has a profound meaning because it reveals
that, for this long-range system, the order in which one
performs the $t \to \infty$ with respect to the $N \to \infty$
limit is crucial. Statistical mechanics describes the situation in
which the infinite time limit is considered {\it before} the
number of particles tends to infinity. Here, it becomes apparent
that reversing the order of the limits leads to a {\it different
equilibrium state}. This aspect has been particularly emphasized
in several papers by C. Tsallis and co-workers (including
Ref.~\cite{lrt2001}), who propose that such a state should be
described by Tsallis statistics~\cite{Tsallisjsp}. However,
because of the absence of derivations from first principles and of
testable predictions of this theory, this cannot be considered as
a satisfactory solution of this puzzling dynamical behaviour. 

Let us remark that, although the description at the beginning of last
paragraph" clarifies the origin of the observed "dynamical" ensemble
inequivalence, it does not give any hint on how to characterize the
system in the QSS state. In the next Section we will show that a
theory based on the Vlasov equation associated to the HMF model
provides fully justified arguments and predictions on the behaviour of
the QSS state.

Let us finally observe that the slow time evolution towards the
Boltzmann-Gibbs equilibrium shown in Fig.~\ref{figdiffN} also
explains why the microcanonical simulations reported in
Fig.~\ref{figcaloric} do not correspond, as expected, to the
theoretical results given by microcanonical and canonical
statistical mechanics. The simulation time was simply too short
and would one have waited longer, the disagreement would have
totally disappeared. Moreover, simulation points corresponding to
initial velocity distributions which have smaller stability
thresholds (see next Section) show a stronger disagreement with
respect to the canonical caloric curve.

\section{The Vlasov equation}

For mean-field systems, and Hamiltonian~(\ref{hamilHMF}) is one
example, it has been mathematically proven~\cite{BraunHepp,Spohn} that,
for a finite time and in the limit $N \to \infty$,
the $N$-particle dynamics is well described by the Vlasov equation.
Let us show how a simple minded kinetic theory allows us to
derive the Vlasov equation.

The state of the $N$-particles
system can be exactly described by
the {\em discrete} single particle time-dependent density
function, whose dynamics is exactly given by the Klimontovich
equation~\cite{fredthierPRE}. However, it is far too precise for
the description we are interested in, since it is a function of
the 2$N$ Lagrangian coordinates of the particles, $\theta_i$ and
$p_i$.

As we are interested in systems with large number of particles,
$1/N$ is a small parameter. This suggests to describe the system
with an asymptotic expansion and to approximate the discrete density
by a continuous distribution $f_0(t,\theta,p)$,
depending on time $t$ and
only on the Eulerian coordinates of the phase space, $\theta$ and $p$.
These steps are explicitly given in Ref.~\cite{nicholson}.
However, what is important for the purpose of this paper, is that
at the lowest order, one gets the Vlasov equation
\begin{equation}
\frac{\partial f_0}{\partial t}+p\frac{\partial f_0}{\partial
\theta} -\frac{{\rm d} \langle V\rangle}{{\rm d} \theta}
\frac{\partial f_0}{\partial p}=0,\label{Vlasov}
\end{equation}
where one has introduced  the averaged potential
\begin{equation}
\langle V\rangle=-\int_0^{2\pi}{\rm d}
  \alpha\int_{-\infty}^{+\infty}{\rm d} p\
  \cos(\theta-\alpha)\, f_0(t,\alpha,p).
  \end{equation}
The right-hand-side of Eq.~(\ref{Vlasov}) is zero because of the
$N \to \infty$ limit. It would be non-zero only if ``collisional"
effects are taken into account. It is however important to
underline that there are no collisions here: granular effects or
finite $N$ corrections would be more appropriate names.

For homogeneous distributions with respect to $\theta$, one gets
$\langle V\rangle=0$. The single particle distribution
$f_{0}(t,p)$ is thus stationary since Eq.~(\ref{Vlasov}) can be
rewritten as
\begin{equation}
  \frac{\partial f_0}{\partial t}(t,p)=0.
  \label{Vlasovhomogeneous}
\end{equation}
This explains the {\it stationarity} property of any homogeneous
distribution $f_{0}(p)$. However, this does not ensure {\it stability}. Two
different methods were subsequently introduced to determine
stability. The first one relies on Lyapunov functional
stability analysis using the energy-Casimir method~\cite{yama},
while the second one considers the poles of the dieletric
constant of the Hamiltonian~(see Ref.~\cite{fredthierPRE,konishi-kaneko}).
In both cases,
one obtains that the homogeneous distribution $f_0(p)$ is stable if
and only if the quantity
\begin{equation}
I=1+\frac{1}{2}\int_{-\infty}^{+\infty}\frac{f_0'(p)}{p}\ {\rm d}
p\label{therholdcon}
\end{equation}
is positive. This condition reveals that there can be an
infinite number of Vlasov stable distributions. Let us briefly
discuss some examples.

\begin{figure}[htb]
\begin{center}
\includegraphics[height=6truecm]{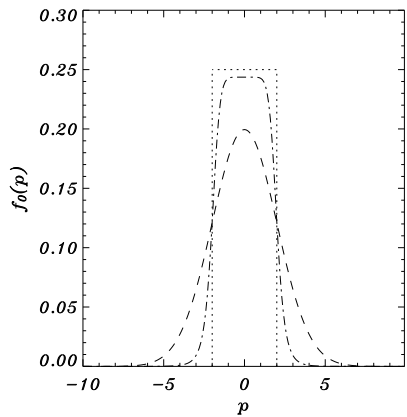}
\end{center}
\caption{Three examples of stationary homogenous solutions of the
Vlasov equation. The Gaussian (dashed), the water-bag (dotted) and
the power-law (Eq.~\ref{eq:powertail}) in the case $\nu=8$
(dash-dotted).} \label{figdiffzero}
\end{figure}

\begin{itemize}
\item The first one is the Gaussian distributions $f_0(p)\sim
\exp({-\beta p^2/2})$ (see Fig.~\ref{figdiffzero}) which is expected
at equilibrium. With the threshold condition~(\ref{therholdcon}), one
recovers the statistical mechanics result that the critical inverse
temperature is $\beta_c=2$, and its associated critical energy
$U_c=3/4$ as plotted in Fig.~\ref{figcaloric}.

\item The
second example is the water-bag distribution, also depicted in
Fig.~\ref{figdiffzero}, which has been often used in the past
to numerically test the out-of-equilibrium properties of the HMF model.
In that case, one obtains a smaller critical
energy $U_c=7/12$.

\item Another example would be the $q$-exponentials of  Tsallis
statistics: $f_0(p)\sim [1-\alpha(1-q)p^2]^{\frac{1}{1-q}}$. In
that case, one gets~\cite{Vallejos} that
$U_c=\frac{3}{4}+\frac{q-1}{{2(5-3q)}}$, recovering the Gaussian
result for $q=1$ and the water-bag one when $q$ approaches
infinity with a cut-off to keep energy finite.

\item The last example is a distribution with power-law tails
\begin{equation}
\label{eq:powertail} f_{0}(p) = \frac{A}{1+|p/p_{0}|^{\nu}},
\end{equation}
where $p_{0} = \sqrt{\frac{\sin(3\pi/\nu)}{\sin(\pi/\nu)}
\frac{K}{N}}$ controls the kinetic energy density $K/N$ and
$A=\nu\sin(\pi/\nu)/(2\pi p_{0})$ is the normalization factor. The
exponent $\nu$ must be greater than $3$ to get a finite kinetic
energy: we have used $\nu=8$ (see Fig.~\ref{figdiffzero}).  Note
that the power-law distribution cannot be included in the
$q$-exponential family, although it has similar power law tails at
large $|p|$. Distribution~(\ref{eq:powertail}) is stable above the
critical energy
$U_{c}=\frac{1}{2}+\frac{\sin(\pi/\nu)}{4\sin(3\pi/\nu)}$.
\end{itemize}

All above distributions are thus stationary solutions of the
Vlasov equation~(\ref{Vlasov}). They are stable provided the
quantity~(\ref{therholdcon}) is positive. However, it is important
to realize that they are Vlasov stable stationary solutions among
{\em infinitely} many others and there is no reason to emphasize
one more than the other.

The existence of an infinite number of Vlasov stable
distributions is the key point to explain the out-of-equilibrium
QSS observed in the HMF dynamics and shown in Fig.~\ref{figdiffN}.
Although we start initially from such a stable state (the
homogeneous water-bag), finite $N$ effects drive the system away
from it, through other stable stationary states. This slow
quasi-stationary evolution across the  infinite number of
stationary and stable Vlasov states finishes with the ultimate
evolution towards the Boltzmann-Gibbs equilibrium state. For the
HMF model, it has been proven~\cite{fredthierPRE} that Vlasov
stable homogeneous distribution functions do not evolve on time
scales of order smaller or equal to $N$. This is in agreement with
the $N^{1.7}$ scaling law numerically found~\cite{yama} for the
relaxation towards the Boltzmann-Gibbs equilibrium state.

\begin{figure}
\begin{center}
{\setlength{\unitlength}{0.5pt}
\begin{picture}(200,480)(10,20)
\put(0,450){\framebox(280,40){Initial Condition}}
\put(0,250){\framebox(280,40){Vlasov's Equilibrium}}
\put(0,50){\framebox(280,40){Boltzmann's Equilibrium}}

\put(175,355){$\tau_v=N^0$}

\put(175,155){$\tau_c=N^\delta$}

\put(25,370){{\em Violent} }

\put(15,340){ relaxation}

\put(5,170){{\em Collisional}}

\put(10,140){ relaxation}

\put(140,230){\vector(0,-1){130}}
\put(140,430){\vector(0,-1){130}}
\end{picture}}
\end{center}
\caption{Schematic description of the typical dynamical evolution
of systems with long-range interactions.}
\label{schematicdescription}
\end{figure}
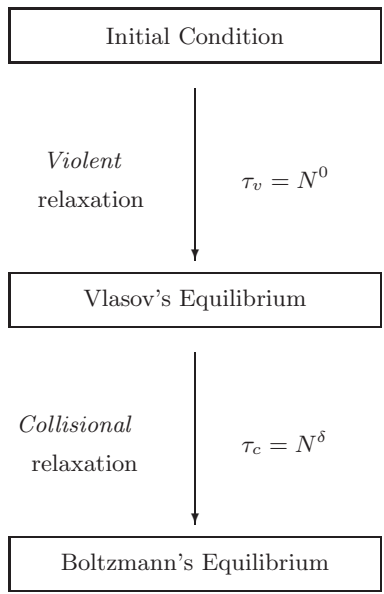

Let us stress that the above scenario is consistent with what
happens generically for systems with long-range
interactions~\cite{Lyndenwood68,ChavanisHouches}.
In a first stage, called {\it violent relaxation} the system goes
from a generic initial condition, which is not necessarily Vlasov
stable, towards a Vlasov stable state. This is a fast process
happening usually on a fast timescale, {\em independent} of the
number of particles. In a second stage, named {\it collisional
relaxation}, finite $N$ effects come into play and the Vlasov
description is no more valid for the discrete systems. The
timescale of this second process is strongly dependent on $N$. One
generally considers that it is a power law $N^\delta$. A typical
example is  Chandrasekhar relaxation time scale for stellar
systems, which is proportional to $N/\ln N$. This scenario of the
typical evolution of long-range systems is summarized in
Fig.~\ref{schematicdescription}.

It is important to remark that, recently, Caglioti and
Rousset~\cite{CagliottiRousset} rigorously proved that for a wide
class of potentials, particles starting close to a Vlasov stable
distribution remain close to it for times that scale at least like
$N^{1/8}$: this result is consistent with the power law
conjectured for collisional relaxation. Unfortunately, apart from
a recent progress ~\cite{haurayjabin}, very few rigorous results
exist in the case of singular potentials, which would be of
paramount importance for Coulomb and gravitational interactions.

\section{Conclusions}

We have emphasized that the slow dynamical evolution of the Hamiltonian
Mean-Field model (HMF) can be well understood with the help
of the Vlasov equation.

Quasi-stationary states (QSS) observed in the $N$-particle dynamics
of the HMF Hamiltonian are nothing but Vlasov stable stationary
states, which lose their stability because of {\it collisional},
finite $N$, effects.

There is an {\it infinity} of Vlasov stable homogeneous (zero
magnetization) states corresponding to different initial velocity
distributions $f_0(t=0,p)$. Taking three examples (Gaussian, water-bag
and power-law), we have shown that their stability domain in
energy is different.

Also Tsallis $q$-exponentials in momentum, homogeneous in angle,
distribution functions are Vlasov stable stationary states in a certain energy
region where QSS are observed in the HMF model. However, they
are not special in any respect, among an {\it infinity} of others.

In the finite $N$ HMF systems, all of them converge sooner or
later to the Boltzmann-Gibbs equilibrium. However, the relaxation
time is shown numerically to diverge with a power-law $N^\delta$,
with $\delta\simeq 1.7$ for the homogeneous water-bag state.
Analytically, one can prove that such a divergence must have
$\delta>1$.

On the time scale $\tau = t/N$ the QSS of the HMF model do not
evolve. However, one can prove~\cite{fredthierPRE} that this is a
peculiarity of one-dimensional models. This time scale is the
appropriate one to study momentum autocorrelation functions and
diffusion in angle. Such issues are discussed in more detail in
Ref.~\cite{fredthierPRE}, where {\it weak} or {\it strong
anomalous diffusion} for angles is predicted, both at equilibrium
and for QSS.

\end{document}